\documentclass[conference]{IEEEtran}
\usepackage{ucs}
\usepackage[utf8x]{inputenc}

\usepackage{cite}
\usepackage{amsmath,amssymb,amsfonts}
\usepackage{algorithmic}
\usepackage{graphicx}
\usepackage{textcomp}
\usepackage{xcolor}
\usepackage{algorithm}

\title{BGP Typo: A Longitudinal Study and Remedies}
\author{\IEEEauthorblockN{Liron David}
\IEEEauthorblockA{Dept.\ of Computer Science and Applied Mathematics \\
Weizmann Institute of Science\\
Email: lirondavid@gamil.com }
\and
\IEEEauthorblockN{Yuval Shavitt}
\IEEEauthorblockA{School of Electrical Engineering,\\
Tel-Aviv University\\
Email: shavitt@eng.tau.ac.il}}

\begin{document}

\maketitle

\begin{abstract}
BGP is the protocol that keeps Internet connected.  Operators use it by announcing Address Prefixes (APs), namely IP address blocks, that they own or that they agree to serve as transit for.  BGP enables ISPs to devise complex policies to control what AP announcements to accept (import policy), the route selection, and what AP to announce and to whom (export policy).  In addition, BGP is also used to coarse traffic engineering for incoming traffic via the prepend mechanism.  

However, there are no wide-spread good tools for managing BGP and much of the complex configuration is done by home-brewed scripts or simply by manually configuring router with bare-bone terminal interface.  This process generates many configuration mistakes. 
 
In this study, we examine typos that propagates in BGP announcements and can be found in many of the public databases. We classify them and quantify their presence, and surprisingly found tens of ASNs and hundreds of APs affected by typos on any given time. In addition, we suggest a simple algorithm that can detect (and clean) most of them with almost no false positives.
 

\end{abstract}

\section{Introduction}

BGP (together with IP) is the glue that keeps the Internet connected.
Surprisingly, although its complexity, BGP is still largely managed manually, and as a result BGP suffers from many configuration errors  \cite{mahajan2002understanding,Feamster-misconf,BGPlies,gilad2018perfect,CAIDA_Hijacking:19}.  One specific problem in BGP announcements is typos, which are inserted by operators, that result is corrupted route announcement. This problem is missing from almost all previous analysis of BGP \cite{BGPanomaliesSurvey}.

Many of the typos in BGP are due to route prepending.  Route prepending is the duplication of an ASN in the route to make it look longer and thus less favorite for routing \cite{Chang-prepending}.  For example, if an AS X has two upstream providers: a main provider M, and a more expensive backup provider B, it may prepend its announcement to B in an attempt to make routes through B less favourable. Namely it will send M the route "X", and B the route "X X X". If both B and M are customers of AS Y, Y will receive the route "M X" from M and the route "B X X X" from B. If no other preference exist the route to M will be selected by Y since it is shorter (2 hops vs.\ 4 hops). Note that the more times X duplicates itself in the route, the higher is the tendency of traffic to shift from B to M \cite{quoitin2005performance}. Note that most prepending are small with 2, 3, and 4 ASN copies, but much larger number are also seen.

Since prepending is done manually, typing mistakes happen with the ASN, such as missing a digit, adding a digit, and swapping digits.  One other common typing mistake is typing the number of prepends at the end of the route, sometimes instead of the prepending and sometimes in addition to \cite{CAIDA_Hijacking:19}. 

Unlike many configuration errors, typing mistakes seem to be benign since they seem not to change the routing, but they effect routing in several ways. If a BGP message arrives at an AS with its ASN in it due to a typo, it will most likely be rejected, since ASes are responsible to check that they do not create AS-level routing loops. This may result in black-holes or longer routes.  The opposite effect may happen if prepending is replaced with typing the number of prepends.  If the original traffic engineering intent was the route "X X X X" we may get instead "X 3" a shorter route and thus less traffic shifting.

The recent increased deployment of RPKI \cite{RPKI} exposes messages with a typo in the first AS vulnerable for special treatment since the 'origin AS' does not match the RPKI record. Depending on the RPKI implementation, such messages will either be ignored or get the lowest selection priority. This may result with black holes, or route length inflation. Roughly half of the typos we found are exposed to this problem.

Finally, erroneous ASN in the path may trigger hijack detection systems \cite{CAIDA_Hijacking:19}, especially in the near future where deep learning approaches will be put into usage \cite{AP2Vec}. Commercial BGP monitoring services are currently reporting a change in the origin AS and upstream provider, thus, almost all typos we find will trigger an alert.  Other hijack detection services perform analysis of the entire route, e.g., by using valley-free analysis \cite{Gao-VF}, making even the less common cases, where the typos are found further left in the route, trigger an alert, as well.

\begin{figure}[htb]
\centering
\includegraphics[width=0.95\linewidth]{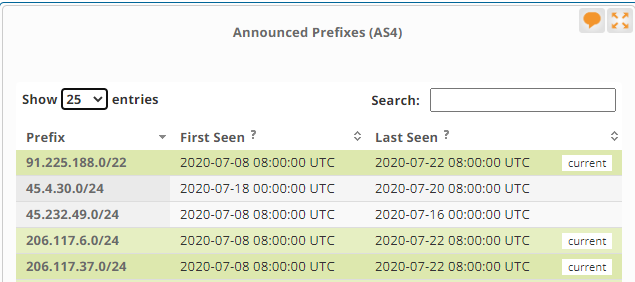}
\caption{Erroneous APs (91.225.188.0/22, 45.4.30.0/24, 45.232.49.0/24) associated with USC (AS4) from RIPE site.}
\label{fig:RIPE-USC-typo}
\end{figure}

\begin{figure}[htb]
\centering
\includegraphics[width=\linewidth]{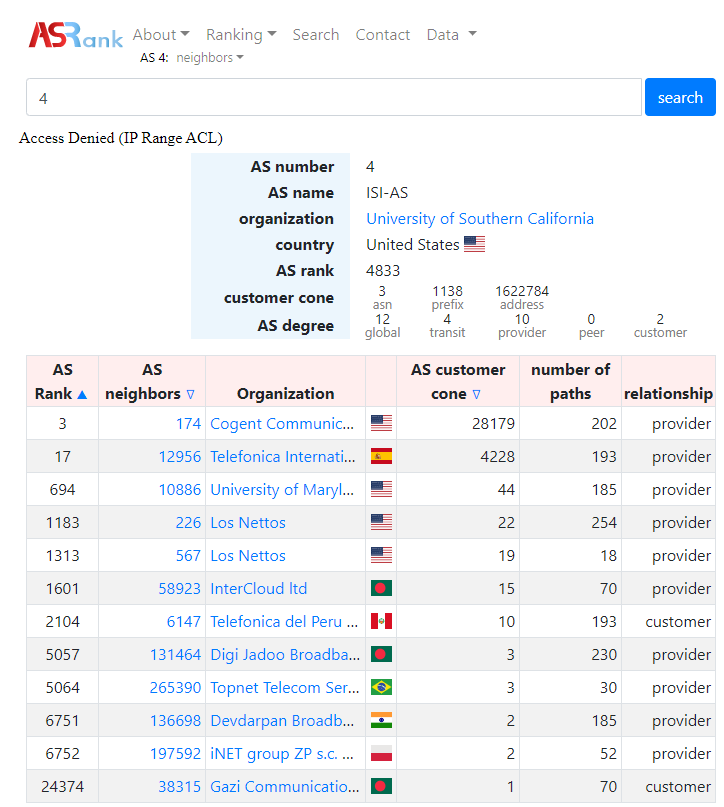}
\caption{Erroneous neighboring ASNs (24374, 6752,6751, 5064, 5057, 2104, 1601) associated with USC (AS4) from CAIDA site.}
\label{fig:CAIDA-USC-typo}
\end{figure}

\begin{figure}[htb]
\centering
\includegraphics[width=0.95\linewidth]{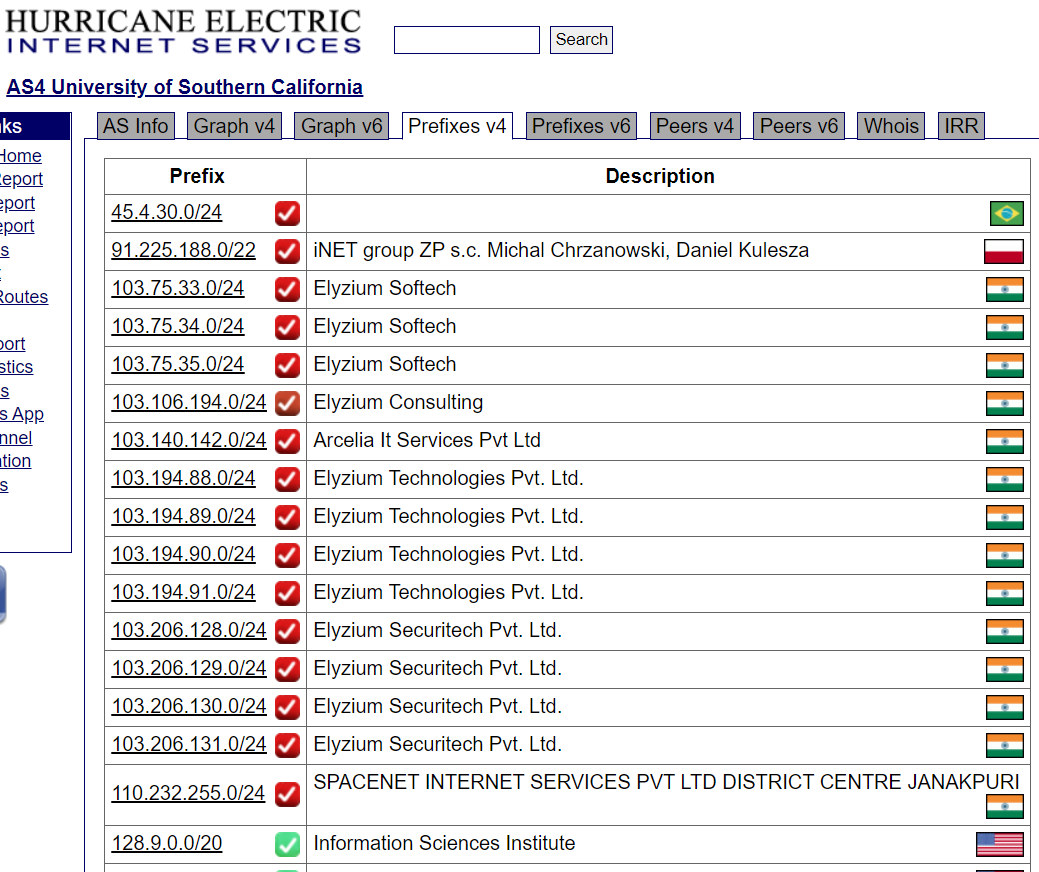}
\caption{Erroneous APs (all except for the last one in the figure are due to typos.  The full list contains additional valid and invalid APs.) associated with USC (AS4) from Hurricane Electric web site.}
\label{fig:HE-USC-typo}
\end{figure}

Today's typos are largely ignored, and as a result they contaminate many public databases: The RIPE database reports wrong AP ownership (Figure~\ref{fig:RIPE-USC-typo}), CAIDA database reports wrong neighborhood between ASNs (Figure \ref{fig:CAIDA-USC-typo}) and deduce their non-existent Type of Relationships in their ToR database, Hurricane Electric (Figure~\ref{fig:HE-USC-typo}) reports wrong AP ownership, and so do other public commercial databases.  We show that this errors can be detected with high accuracy and almost with no false detection and removed from BGP, as well as from these databases. 

The rest of the paper is organized as follow.  In section \ref{sec:methods} we describe the different and classify the typos and describe the data collection and analysis. Section~\ref{sec:long} brings the longitudinal (and other related) analysis.  Section~\ref{sec:detection} discusses typo detection, and final section discusses the results.


\section{Methods and Data} \label{sec:methods}

\subsection{Typo classification} \label{subsec:classification}

We classified the typing errors to the following categories:
\begin{enumerate}
    \item {\bf prepend count} -- In this scenario, the operator inserts the prepend amount before the ASN, so if one wants to prepend AS X three times the results will be either "X 3"  or "X X X 3".  As a result, ASes with with small ASNs (AS3 which belongs to MIT in this example) are seen as the co-owners of many blocks that are not connected to them.
    
    For example, from late 2016 till early 2020, 202.56.168.0/23, which is owned by AS131758 is prepended twice and routes end with the AS sequence: "18351 131758 131758 131758".  At the same time 202.56.168.0/24, is announced with the route ending by "18351 131758 3", obviously an attempt to prepend thrice (Figure~\ref{fig:AS131758} depicts the announcements).  Needless to say, AS3 (MIT) has no relationships with AS131758 (PT.\ Bali Ning, a small ISP from Bali, Indonesia). Interestingly, comparing all the 38 routes for the two APs in a one month example, there was no difference in routing in any of them, albeit the difference of 1 in the AS path length.
    
\begin{figure}[htb]
\centering
\includegraphics[width=0.95\linewidth]{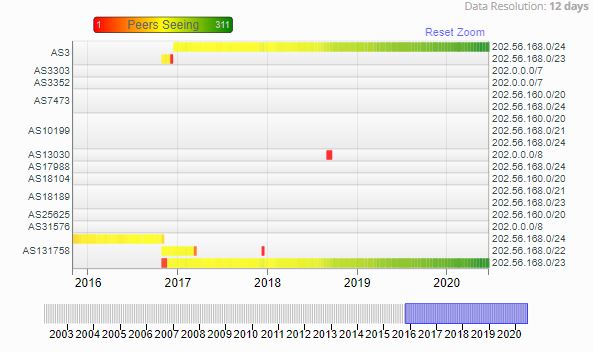}
\caption{Announcement history of 202.56.168.0/24 from the RIPE site.}
\label{fig:AS131758}
\end{figure}
    
    In the following examples, the number of prepends at the end of the path corresponds to the number of repetitions (prepending) of the ASN.
    \begin{itemize}
       \item 5.100.240.0/24 ``7660 2516 701 3356 42020 42020 42020 42020 42020 42020 31126 56902 56902 56902 3"
       \item 5.183.170.0/24 ``3549 3356 8708 213323 213323 213323 213323 213323 213323 213323 213323 8"
       \item 103.49.56.0/24 "57463 9498 133718 133718 2"
    \end{itemize}
    
    We searched for cases where the prepend number is in the range [1,12], though there are also rare examples where we found 15, for example this route from February 25, 2020: "95.130.39.0/24     87.121.64.4              0      0      0 57463 6663 9002 2588 198138 198138 198138 198138 198138 198138 198138 198138 198138 198138 198138 198138 198138 198138 198138 15 ".

    \item {\bf swap digits} - In this scenario, the operator, when prepending an ASN, swaps two consecutive digits. 
    In this example from December 2019 for the AP 2.58.168.0/22 we see the route "3741 9002 29278 {\em 29728} 29278 29278 29278 29278 29278 42864 i"
    
    \item {\bf insert/delete one digit} - In this scenario, the operator, when prepending the AS, inserts or deletes one digit from the original ASN.
    In this example from November 2019 for the AP 81.88.208.0/20  "24441 6939 20764 39709 39709 {\em 3970} 39709 39709 i"
    
    \item {\bf missing space } - In this scenario, the operator, when prepending an AS, misses a space between two ASNs. For example, in April 2017 the AP 72.5.214.0/24  announcement appears as  "3356 32026 32026 32026 32026 32026 32026 {\em 3202632026} 32026 32026 32026 i"

    \item {\bf Other typos} - We found rare typos that were not counted like:
    \begin{itemize}
    \item type 1 typos with very large numbers such as 15 or 20, for example this route from February 25, 2020: "95.130.39.0/24     87.121.64.4              0      0      0 57463 6663 9002 2588 198138 198138 198138 198138 198138 198138 198138 198138 198138 198138 198138 198138 198138 198138 198138 15".
    
    \item private ASNs that may be a leak but looks suspiciously similar to the ASN next to them, for example: "65100 100" and "65292 292".
    \end{itemize}

\end{enumerate}

\subsection{Data Collection and Identifying Typo} \label{sec:data}

The data we examined was from the RouteViews project~\cite{RV}, where we extract one RIB file from the first day of every month. We run a script that collected the suspected announcements for each typo type. From each suspected announcement, we extracted the suspected ASN pairs, and each pair was manually evaluated to determine if it is legitimate.  We ended up with three categories: typo, OK, or unknown; the latter was quite small since it was quite easy in most cases to determine the legitimacy of a path.

One exception to the above process was type 1 mistakes, since the problem there is limited to identify the legitimate upstream providers of 12 stub ASNs (1-12).  After examining three months of data, we discovered all the legitimate upstream providers, and any other provider was assumed a typo.

To identify the suspected routes with typo of type 2-4 we used the following procedures ($||$ denotes concatenation):

\begin{algorithm}
\begin{algorithmic}[htb] 
\renewcommand{\algorithmicrequire}{\textbf{Input:}}
\renewcommand{\algorithmicensure}{\textbf{Output:}}
\REQUIRE{$AS_1,AS_2$}
\RETURN $len(AS_1)==len(AS_2)$ and $ \exists i$ s.t.	$AS_1[i]==AS_2[i+1]$ and $AS_1[i+1]==AS_2[i]$ and $\forall j \neq i,i+1$ $AS_1[j]==AS_2[j]$
\caption{swap}
\end{algorithmic}
\end{algorithm}

\begin{algorithm}
\begin{algorithmic}[htb] 
\renewcommand{\algorithmicrequire}{\textbf{Input:}}
\renewcommand{\algorithmicensure}{\textbf{Output:}}
\REQUIRE{$AS_1,AS_2$}
\IF {$len(AS_1)<len(AS_2)$}
\STATE $n=len(AS_2)$
\RETURN $ \exists i$ s.t. $AS_1 == AS_2[1:i-1]$  $||$ $AS_2[i+1:n]$
\ELSE
\STATE $n=len(AS_1)$
\RETURN $ \exists i$ s.t. $AS_2 == AS_1[1:i-1]$  $||$ $AS_1[i+1:n]$
\ENDIF
\caption{insert/delete}
\end{algorithmic}
\end{algorithm}

\begin{algorithm}
\begin{algorithmic}[htb] 
\renewcommand{\algorithmicrequire}{\textbf{Input:}}
\renewcommand{\algorithmicensure}{\textbf{Output:}}
\REQUIRE{$AS_1,AS_2$}
\RETURN $AS_1==AS_2||AS_2$ or $ AS_2==AS_1||AS_1$
\caption{missing space}
\end{algorithmic}
\end{algorithm}

    


To manually determine the legitimacy of a path we used several deduction methods. Of course, routes with loops were declared a typo. Most of the problematic pairs were at the beginning of the AS path (the right hand side), thus, if the penultimate ASN owns the AP, the path was declared a typo. Otherwise, there were many cases were a small ISP or stub AS from one country, say the US, was between two ASes from a different country, say Brazil. In some cases the AS types along the route were unreasonable, e.g., the announcement of Rochester Institute of Technology (AS4385) for 129.21.0.0/16 from November 2017, "6939 3356 385 4385 4385 4385 i", puts AS385, which belongs to the US Air Force, as its transit provider, clearly an unreasonable scenario.
Pairs with reasonable chances to have peering relationships, were declared OK.
There were cases where it was not clear if the path at hand is a typo, or a routing glitch. These case were put in the 'unknown' category, as well as cases were the chances of peering between ASes seemed slim but not totally unreasonable.

\section{Longitudinal Study}  \label{sec:long}

\begin{figure*}[p]
\centering
\includegraphics[width=0.95\linewidth]{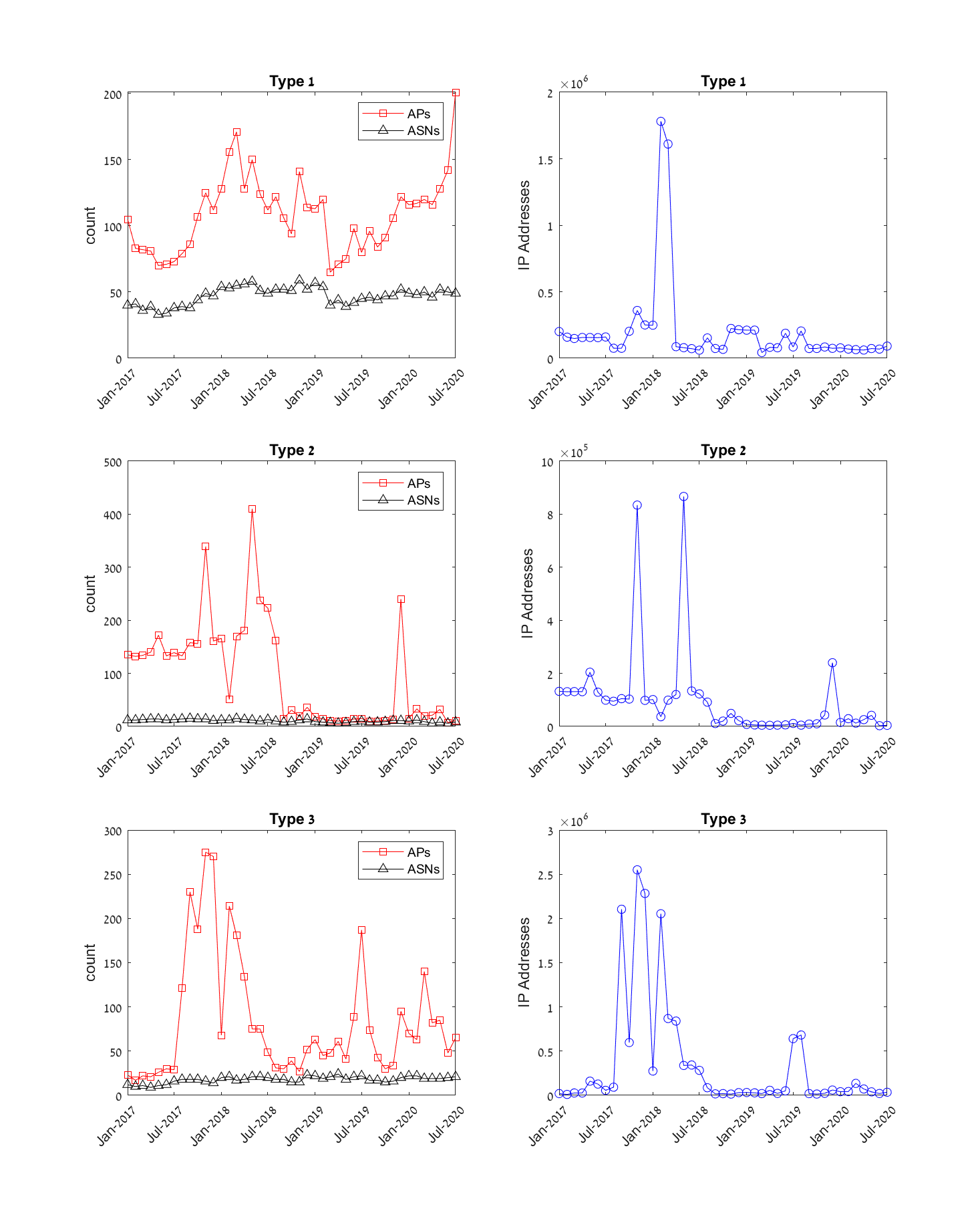}
\caption{The number of APs, ASNs, and IP address affected by type 1 (top), 2 (middle), and 3 (bottom) typos over time.}
\label{fig:resType-all}
\end{figure*}

Figure \ref{fig:resType-all} depicts, for each type of typo and for each month from January 2017 till July 2020, the number of IP addresses in which there are typos, the number of APs for which there are typos, and the number of ASN pairs that comprise a typo. For type 1, instead of typo pairs, we count the number of distinct ASNs, which come right before (form left side) the prepend digit. The figure shows that there is no tendency of the problem to improve nor to worsen for type 1 typos, but there is a clear decline in the number of affect APs for type 3 and even more so for type 2 typos. 

The figures of the number of IP addresses affected by typos show occasional spikes.  We examine the spike in February-March 2018 of the type 1 graph (top-right) and it is due to Belgian Proximus (AS5432) publishing five /14 APs with a typo.  
The bottom row in Figure~\ref{fig:resType-all}  shows a peak  in type 3 typos in January 2017. We manually verified this month and see that in January 2017 2.176.0.0/12 is announced (over a million IP addresses) by Iran's AS12880 with the AS path ``53767 3257 1299 12880 12880 12880 12880 {\em 2880} 12880".  Quite a few  /16 APs that are not part of 2.176.0.0/12, all with the same typo are also announced and together they explain the jump in IP addresses.
The peak in May 2018 for type 2 IP addresses was also manually verified as discussed two paragraphs below.  We are, thus, confident that other peaks in the IP address graphs represent real typos and not due to errors.

Next we discuss the number of APs.
There is a large increase in the number of APs affected by type 1 error in July 2020.  In August (not shown in the figure) this number abates to the levels of May-June 2020.
The second row in Figure~\ref{fig:resType-all}  shows swap errors (type 2).  
Interestingly, in about half of the depicted months a few ASNs (10-15) inserted typos to announcements of 130-410 APs. We manually verified a few of these months to make sure there are no mislabeled pairs. For example, in May 2018 Vietnamese FPT Telecom (AS18403) published the route "37100 6453 18403 {\em 18043} 18403 18403" for 184 APs with 737,280 IP addresses.  TATA (AS6453) is a provider of FPT Telecom, while AS18043 is a stub AS from Taiwan, so this BGP path is certainly an error even without the loop it contains.

We omitted the graph for type 4 typos since there are very few of them.  Throughout the study period, there was, at most, a single problematic ASN with up to 3 affected APs, and up to 3328 affected IP addresses. Only two such ASNs were detected.

\begin{figure}[htb]
\centering
\includegraphics[width=\linewidth]{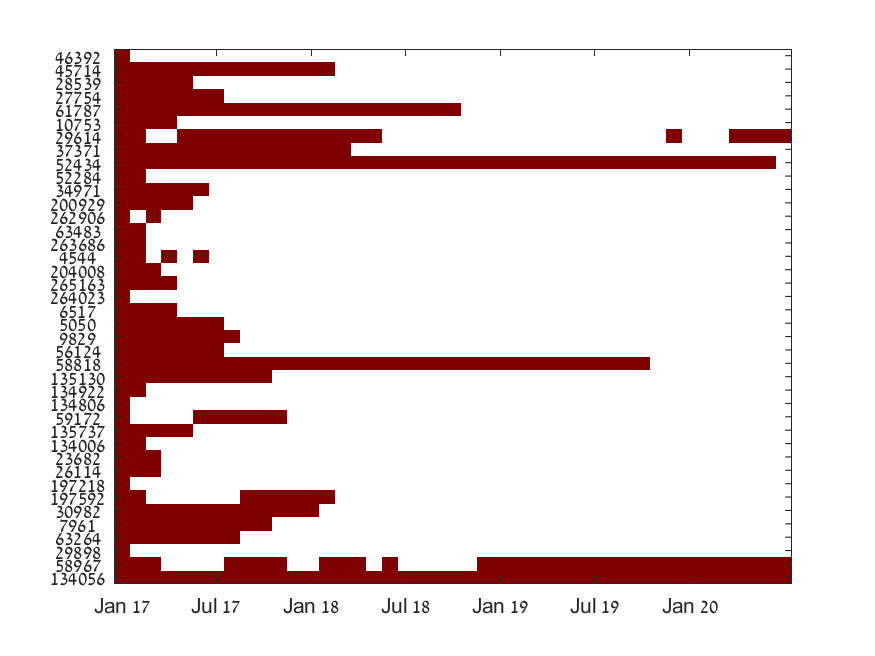}
\caption{Timeline of Jan.\ 2017 ASNs with type 1 typos. A segment is colored if the ASN had a typo in the corresponding month.}
\label{fig:ty1_pairs_time_v2}
\end{figure}

\begin{figure}[htb]
\centering
\includegraphics[width=\linewidth]{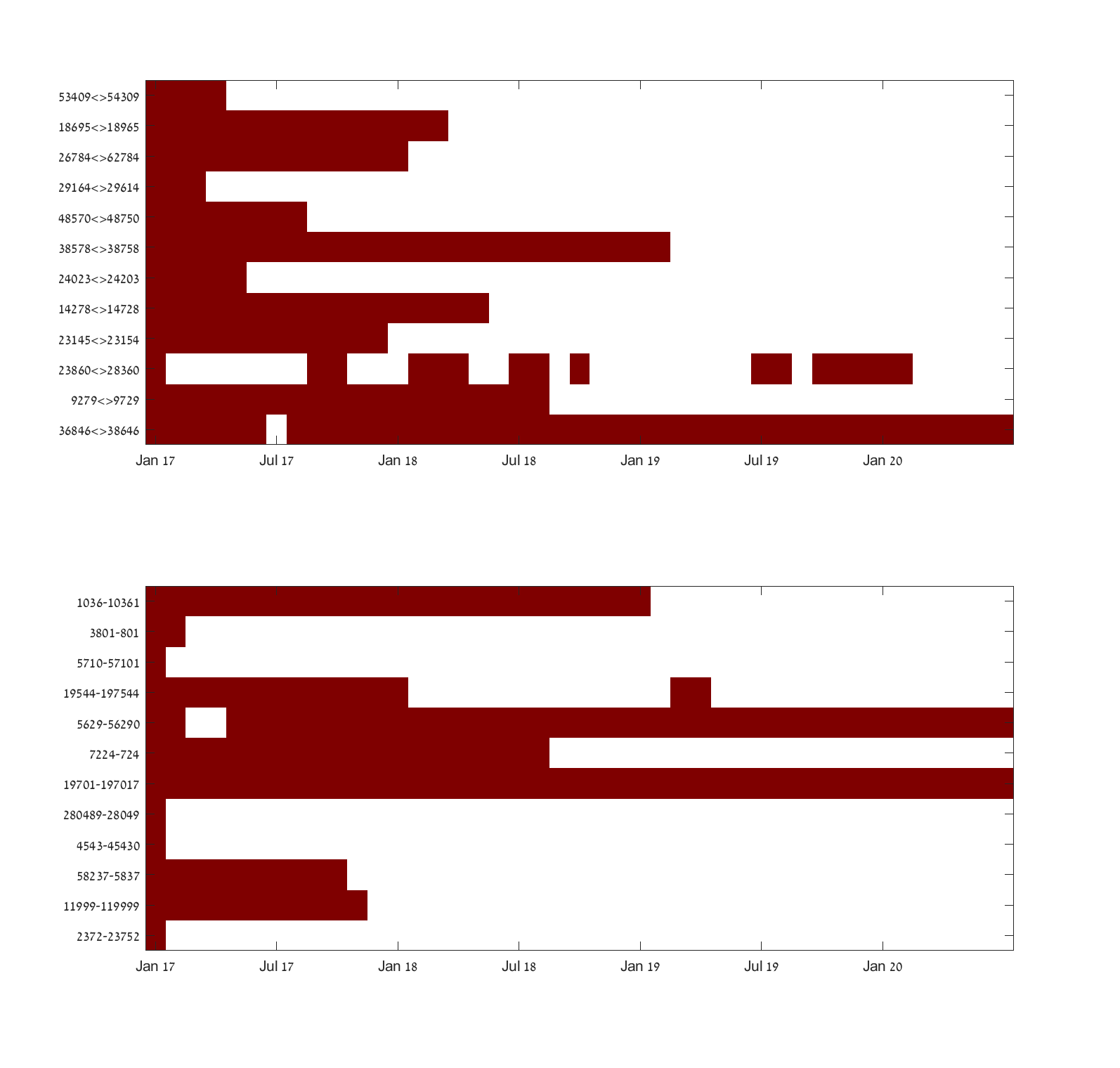}
\caption{Timeline of Jan.\ 2017 ASN pairs with swap typos (top) and insert/delete typos (bottom). A segment is colored if the typo was found in the corresponding month.}
\label{fig:ty2_3_pairs_time}
\end{figure}

Next we would like to study how long typos last.  We searched for the typos found in January 2017, the first month of our study, in all months from February 2017 until July 2020.  We depict the months when they existed in Figures~\ref{fig:ty1_pairs_time_v2} and \ref{fig:ty2_3_pairs_time}. The y-axis of Figure~\ref{fig:ty1_pairs_time_v2} is type 1 errors that appear in November 2017 and in Figure~\ref{fig:ty2_3_pairs_time} is type 2 (top) and 3 (bottom) errors that appear in November 2017. If an error x appears at month y, the rectangle in location (x,y) is colored. Of course, the first column, corresponding to November 2017, is all colored and then most of the typos continue to appear in the following months, some of them  disappear for a while and then occur again. Interestingly, a few typos were seen along the entire study period.

\begin{figure}[htb]
\centering
\includegraphics[width=\linewidth]{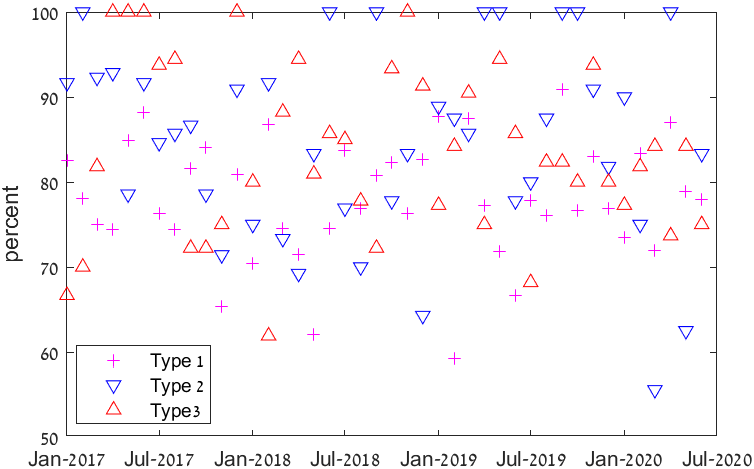}
\caption{The percentage of errors in one month that were seen in the following month.}
\label{fig:errors_m2m}
\end{figure}

To quantify typo persistence, Figure~\ref{fig:errors_m2m} plots the percent of typos in each month that were seen also in the following month, i.e, for each month x from January 2017 to July 2020, we calculate the percentage of errors in month x that appear also in month x+1.  As we can see on Figure~\ref{fig:errors_m2m}, almost all plotted markers are above two thirds, and the averages for type 1, 2 and 3 are 77.9\%, 84.7\% and 83.5\% respectively.
This means we can approximate the typo duration as a Geometric random variable with $p=0.8$. The average duration is thus $1/(1-p)$, namely 5 months.  


\subsection{Mistake Prevalence}

\begin{figure*}[bt]
\centering
\includegraphics[width=0.9\linewidth]{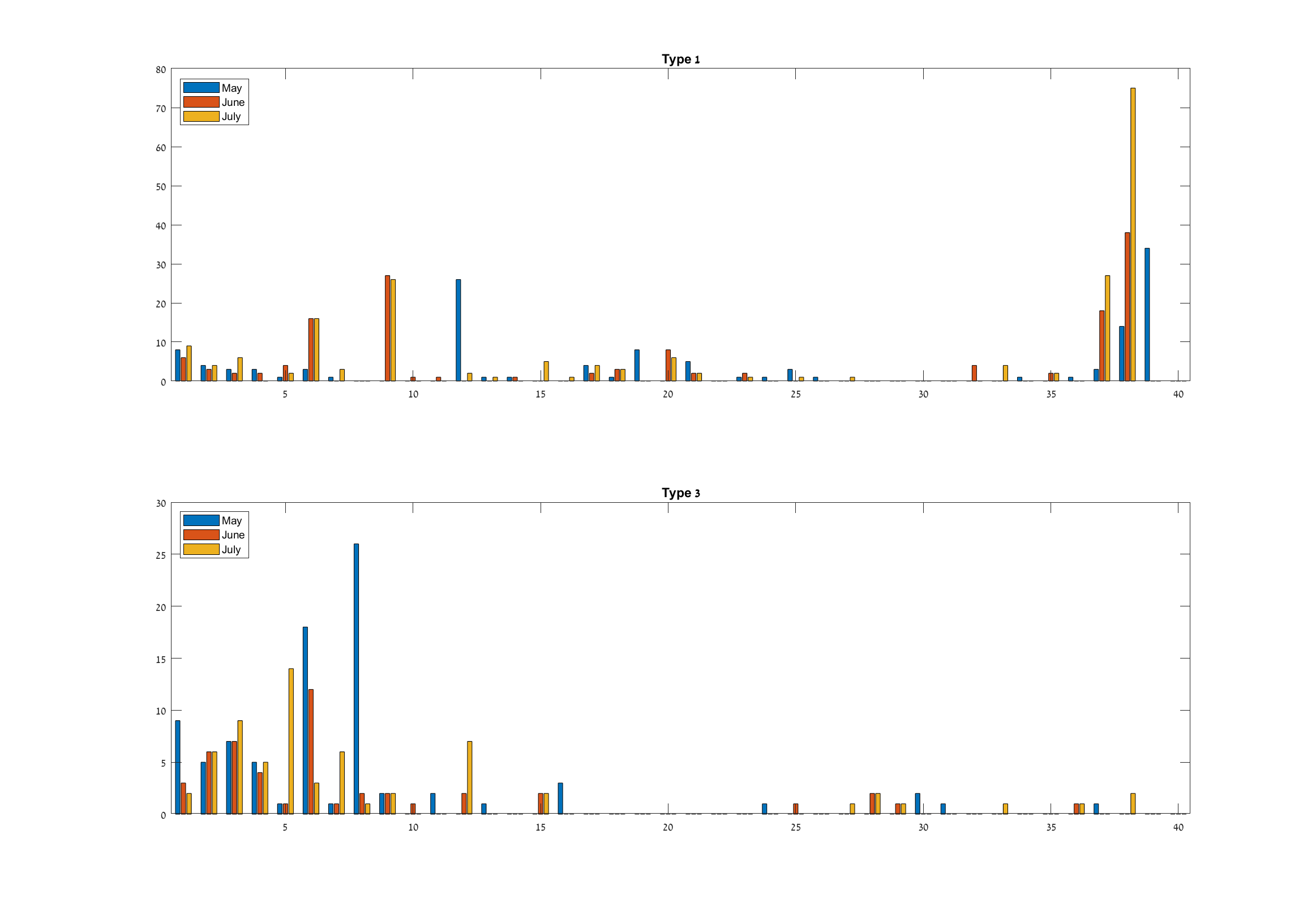}
\caption{A histogram of the number of BGP listener seen each typo in the RouteView files for May-July 2020.}
\label{fig:prevel}
\end{figure*}

Figure \ref{fig:prevel} shows a histogram of the number of BGP listeners (collectors)  seen each typo in the RouteView files for May-July 2020. Only Type 1 and Type 3 typos are presented since there were fewer type 2 typos and they behave similarly to type 3 typos.  Note that any AP announcement is seen by a different number of collectors, all bounded by 41 collectors, and the number of collectors with a typo are different.


Type 3 and 2 typos are usually seen only by less than a ten collectors.  Type 1 mistakes have a significant portions of the typos propagating in to almost all the collectors, 36-38 of them.

\begin{figure*}[htb]
\centering
\includegraphics[width=0.9\linewidth]{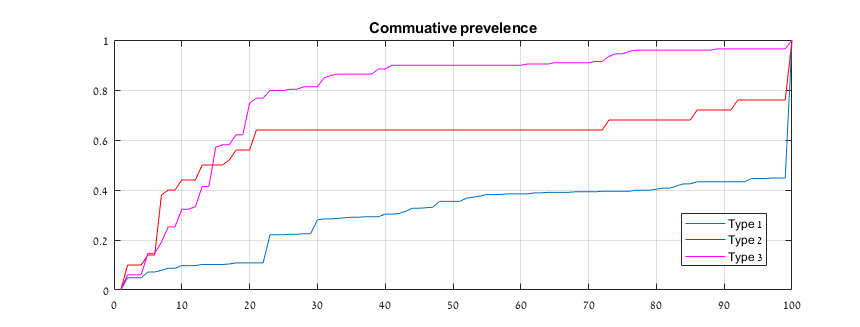}
\caption{The cumulative pairs portion with at least X percent of BGP listeners that saw the typo  in the RouteView files for May-July 2020.}
\label{fig:prevel_pdf}
\end{figure*}

Figure~\ref{fig:prevel_pdf} shows a cdf of the cumulative  percent of mistake seen by RouteViews listeners (X axis) summed by typo for the months May-July 2020.  For type 1 typos over 50\% of the typos are seen at all the BGP listener that have an entry for the relevant AP; for type 2 over 20\% of the typos are seen at all the BGP listener; and for type 3 this number is negligible.  The majority of the typos of type 2 and 3 propagate to less than 20\% of the BGP listener.

\subsection{Offenders Study}

Interestingly, the majority of the type 1 problems come from 4 countries: Brazil, Indonesia, India and the USA,  the first three are responsible for close to half the typos. However, while 28.8\% of the Internet ASNs are registered as American (see Table~\ref{tab:topAS}), only 12\textonequarter\% are registered to the other three combined.

\begin{figure}[htb]
\centering
\includegraphics[width=0.95\linewidth]{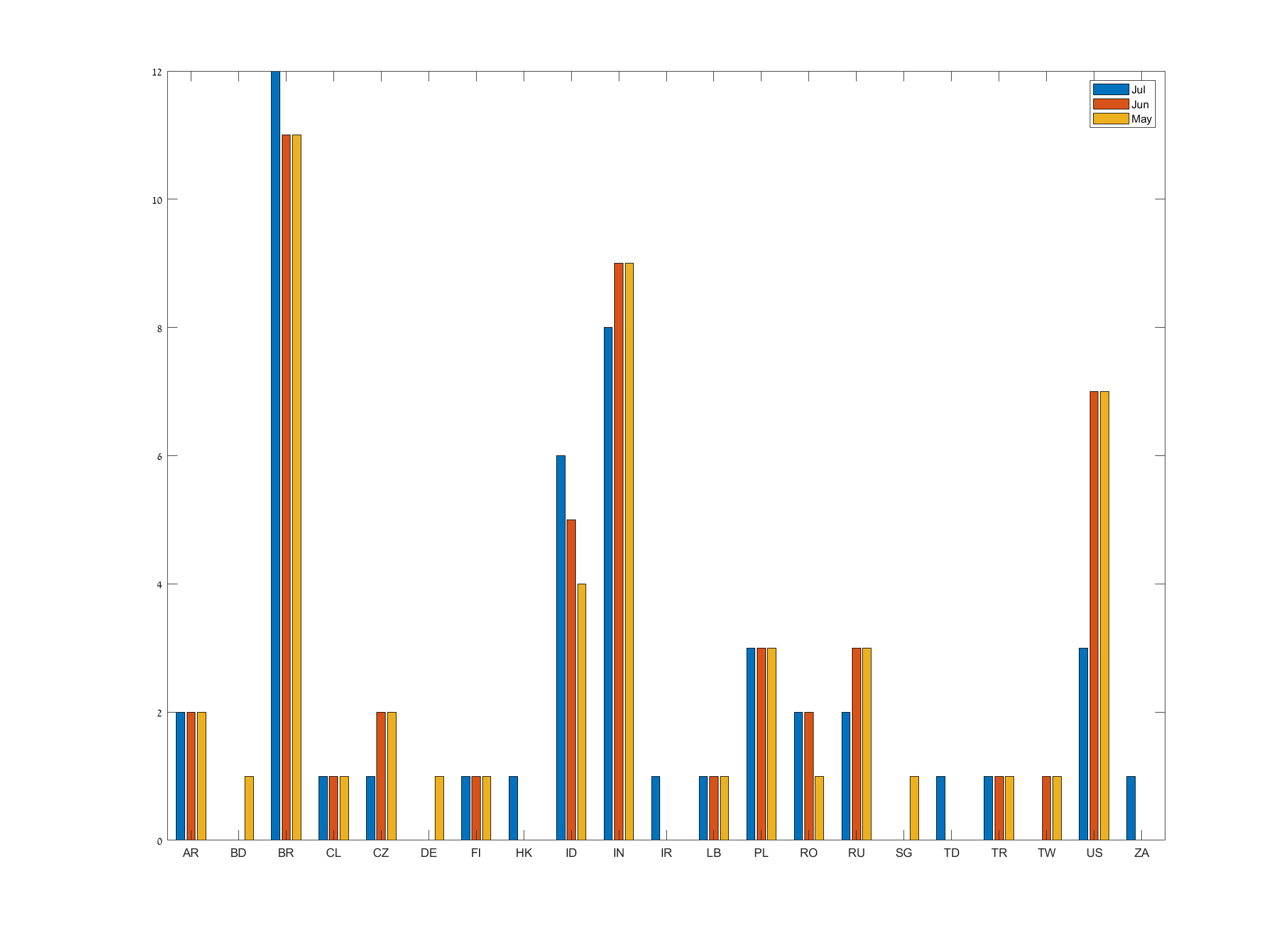}
\caption{Origin country of ASNs with Type 1 typos for May-July 2020.}
\label{fig:t1_country}
\end{figure}

\begin{table}[hb]
    \centering
    \begin{tabular}{|c|r|}
 \hline      
country& \%\\ \hline 
US&28.83\\
BR&7.80\\
RU&6.83\\
GB&3.05\\
DE&2.98\\
AU&2.73\\
PL&2.73\\
IN&2.65\\
UA&2.38\\
CA&2.22\\
ID&1.81   \\\hline 
    \end{tabular}
    \vspace{3pt}
    \caption{Countries with most ASNs registered.}
    \label{tab:topAS}
\end{table}


\section{Typo Detection }  \label{sec:detection}

In this section we present simple procedures to clean BGP tables from typos. We then evaluate their success and false alarm rate.

\subsection {Type 1 detection}

As we mentioned in Sec.~\ref{sec:data}, identifying type 1 mistakes is fairly easy, one only needs to map all the upstream providers of ASNs 1-12, and every route that has an ASN with a small number as its origin followed by an unknown upstream providers can be classified as a typo.  

Upstream providers are fairly static, so examining and updating the list periodically should be a problem (e.g., twice a year examination is sufficient).  Alternatively, one can use a list of IP blocks owned or used by these 12 ASes and flag announcements of APs not contained in these blocks.  The block ownership list for these ASes, unlike the list of APs, is also almost static.

\subsection{Type 2 \& 3 detection}

As we showed in Sec.~\ref{subsec:classification} many routes have cycles and this alone can identify many of the typos.  Thus, we suggest the following procedures to identify a typo
\begin{enumerate}

    \item If the pair is part of a loop then it is a typo. For example, in November 2017 the route to AP 143.208.148.0/24  was ``293 6453 7738 8167 25933 25933 25933 264092 264092 \emph{26402} 264092 i''.
 
    \item If the pair has no ASN to its right, check if the right most ASN (the origin) is the owner of the AP. If  the right most is not the owner, the pair is a typo. For example, in November 2017  the route to the AP 155.133.83.0/24 was ``1221 4637 174 50607 62081 199250 199250 199250 19925 19925 i'' and the owner of 155.133.83.0/24 is AS199250. 
    
    \item Validate that the pair together with its immediate neighboring ASNs on either side do not cause a geographic ownership cycle that span non-neighboring countries.  For example, in the April 2020 announcement of AP 193.93.216.0/22 "37100 37100 3255 49284 49824 49824 49824 49824 49824", the country registration of the triplet (3255 49284 49824) is (UA, IT, UA).
    \item If one ASN is not active, e.g., does not announce any AP, it is a typo.
\end{enumerate} 

Note that testing for ownership of an AP is somewhat problematic.  Many of the available databases, such as the one at RIPE (https://stat.ripe.net/) are built based on AP announcements, so cannot help in our case.  Another problem is announcements by DDoS mitigators and other security services that announce APs occasionally, e.g., when a DDoS attack is detected.  The recent increase in using RPKI \cite{RPKI} can help in this task.

\subsection{Type 4 detection}

In the few cases we found, the ASNs that resulted from a missing space had at least 8 decimal digits, thus were obviously not valid.  Currently, the largest assigned ASN is below 400,000, so we can set the threshold there or with sufficient margins, say at 500,000.

\subsection{Identification Evaluation}

We tested a simple algorithm based on the above procedure to detect typos in routes.  The algorithm accepts as input only paths that were flagged as suspicious type 2 or type 3 typos.
\begin{algorithm}
\begin{algorithmic}[htb]
\IF {ASN cycles detected }
\STATE  typo
\ELSE \IF {origin ASN is false}
\STATE typo
\ELSE \IF {Geo cycles detected}
\STATE typo
\ELSE
\STATE no typo
\ENDIF
\ENDIF
\ENDIF
\caption{Typos detection in routes.}
\end{algorithmic}
\end{algorithm}

\begin{figure*}[p]
\centering
\includegraphics[width=0.99\linewidth]{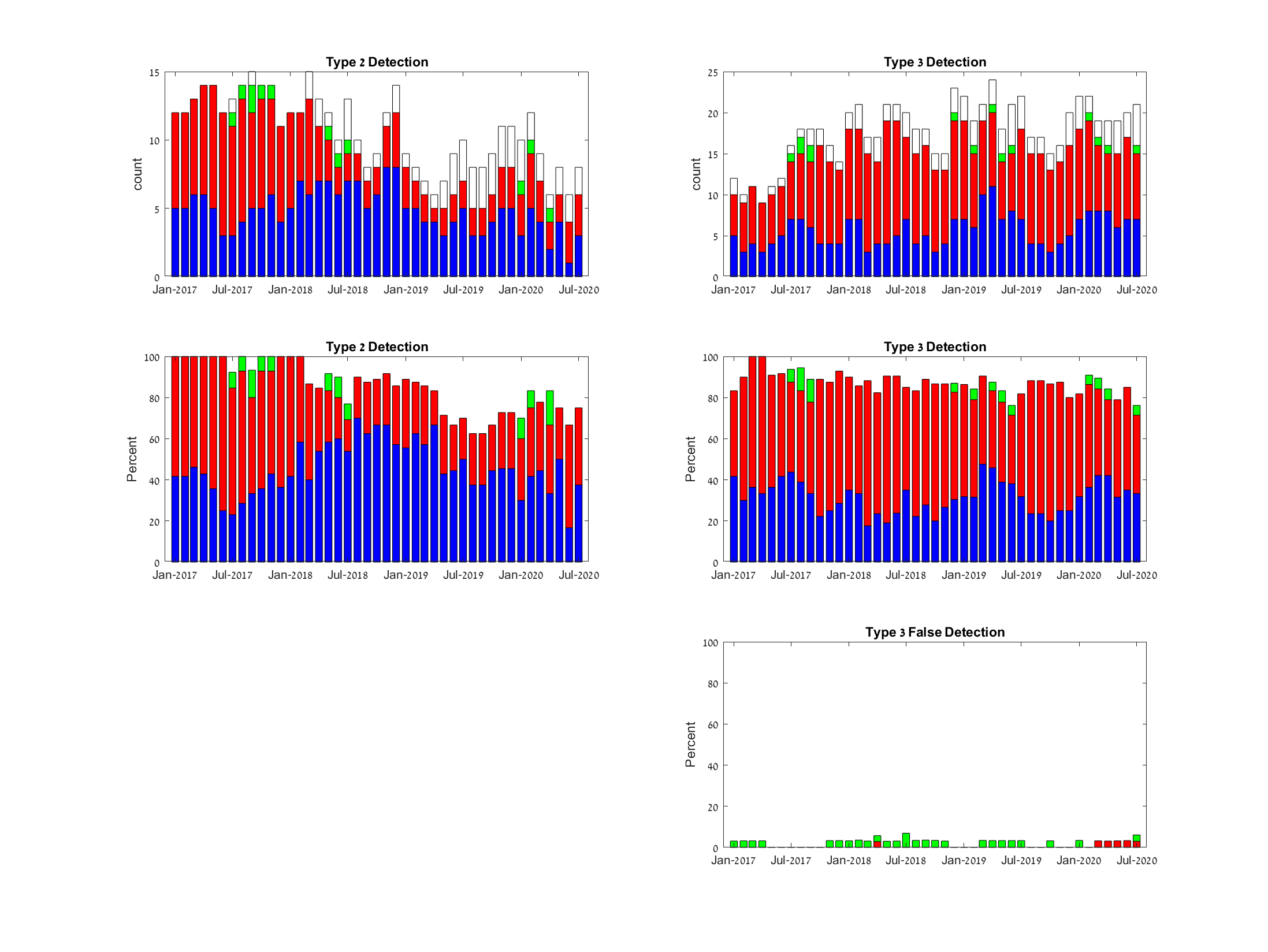}
\caption{Detection success and false detection (bottom row) of type 2 and 3 typos over time.}
\label{fig:detect_FA}
\end{figure*}

Figure \ref{fig:detect_FA} shows the performance of the algorithm for type 2 and 3 typos over time.
The Bars are colored from bottom up according to the stage of the algorithm were detection occurred.
Blue for ASN cycles, red for origin mismatch, and green for geographic loops. The white bar is for undetected pairs.
The typo detection rate is almost always between 80\% and 100\% (middle row in Fig.\ref{fig:detect_FA}) while the number of undetected pairs (white in the top row of Fig~\ref{fig:detect_FA}) is very small.
Interestingly, the detection rate drops with time.

The false detection of routes as typos was quite small.  For type 2 typos we did not have even a single false detection (there were 9-18 such pairs every month).  For type 3 typos the false detection rate was between 0\% and 7\%, or 1-2 false detections out of 28-35 suspicious routes. 
Note that most of the false detection come from the geo-cycles, which contribute very little to the detection.  So at least for type 3 typos one may consider not to use this test in the algorithm.

\section{Discussion}

As we already mentioned, prepending is done manually in many cases. The command line interface enables prepending both by entering the prepend route as it should appear in BGP or by typing the number of times to prepend.  For example the Cisco IOS-XR command definition is:
\begin{verbatim}
set as-path { tag | 
  { prepend as-num [ ... as-num ] | 
  last-as num }}
\end{verbatim}
namely the operetor of AS 1 can prepend by writing "set as-path prepend 1 1 1" or "set as-path prepend last-as 3".   In the older IOS, prepend definition was part of the route-map and required operators to write the prepend route verbosely. 

Juniper (and Huawei most likely), on the other hand, does not support the option to define the number of prepends
\begin{verbatim}
set as-path prepend { list listName | 
  asPathNumber [ asPathNumber ]* }
\end{verbatim}

With these options it is not unlikely that type 1 typo will occur since one can easily forget to remove the number of prepends when entering the full prepend path.  Indeed, in our data, the number of such errors is quite large. 

While we present here ways to detect most of the typos we discussed,  they are only one example of the many BGP configuration errors reported in the literature \cite{mahajan2002understanding,Feamster-misconf,BGPlies}.  Thus, we believe the true solution for BGP configuration is to mechanize this process with better tools.  Recently, the research community start suggesting solutions in this directions \cite{DeepBGP-2020} and we hope that the operator community will follow this direction.

\section*{Acknowledgement}

This research was funded in part by a grant on cyber research from the Israeli PMO and the Blavatnik Interdisciplinary Cyber Research Center at Tel Aviv University.

\bibliographystyle{IEEEtran}
\bibliography{bgp.bib}

\begin{thebibliography}{10}
\providecommand{\url}[1]{#1}
\csname url@samestyle\endcsname
\providecommand{\newblock}{\relax}
\providecommand{\bibinfo}[2]{#2}
\providecommand{\BIBentrySTDinterwordspacing}{\spaceskip=0pt\relax}
\providecommand{\BIBentryALTinterwordstretchfactor}{4}
\providecommand{\BIBentryALTinterwordspacing}{\spaceskip=\fontdimen2\font plus
\BIBentryALTinterwordstretchfactor\fontdimen3\font minus \fontdimen4\font\relax}
\providecommand{\BIBforeignlanguage}[2]{{%
\expandafter\ifx\csname l@#1\endcsname\relax
\typeout{** WARNING: IEEEtran.bst: No hyphenation pattern has been}%
\typeout{** loaded for the language `#1'. Using the pattern for}%
\typeout{** the default language instead.}%
\else
\language=\csname l@#1\endcsname
\fi
#2}}
\providecommand{\BIBdecl}{\relax}
\BIBdecl

\bibitem{mahajan2002understanding}
R.~Mahajan, D.~Wetherall, and T.~Anderson, ``Understanding {BGP} misconfiguration,'' \emph{ACM SIGCOMM Computer Communication Review}, vol.~32, no.~4, pp. 3--16, 2002.

\bibitem{Feamster-misconf}
N.~Feamster and H.~Balakrishnan, ``Detecting {BGP} configuration faults with static analysis,'' in \emph{USENIX NSDI'05}, Boston, Massachusetts, USA, May 2005.

\bibitem{BGPlies}
J.~M. {Del Fiore}, P.~{Merindol}, V.~{Persico}, C.~{Pelsser}, and A.~{Pescapé}, ``Filtering the noise to reveal inter-domain lies,'' in \emph{Network Traffic Measurement and Analysis Conference (TMA)}, 2019, pp. 17--24.

\bibitem{gilad2018perfect}
Y.~Gilad, T.~Hlavacek, A.~Herzberg, M.~Schapira, and H.~Shulman, ``Perfect is the enemy of good: Setting realistic goals for {BGP} security,'' in \emph{The 17th {ACM} Workshop on Hot Topics in Networks}, 2018, pp. 57--63.

\bibitem{CAIDA_Hijacking:19}
S.~Cho, R.~Fontugne, K.~Cho, A.~Dainotti, and P.~Gill, ``{BGP} hijacking classification,'' in \emph{Network Traffic Measurement and Analysis Conference (TMA)}, Jun 2019.

\bibitem{BGPanomaliesSurvey}
B.~Al-Musawi, P.~Branch, and G.~Armitage, ``{BGP} anomaly detection techniques: A survey,'' \emph{{IEEE} Communications Surveys Tutorials}, vol.~19, no.~1, pp. 377--396, 2017.

\bibitem{Chang-prepending}
R.~Chang and M.~Lo, ``Inbound traffic engineering for multihomed {AS}s using {AS} path prepending,'' \emph{IEEE network}, vol.~19, no.~2, pp. 18--25, 2005.

\bibitem{quoitin2005performance}
B.~Quoitin, C.~Pelsser, O.~Bonaventure, and S.~Uhlig, ``A performance evaluation of {BGP}-based traffic engineering,'' \emph{International journal of network management}, vol.~15, no.~3, pp. 177--191, 2005.

\bibitem{RPKI}
G.~Huston and R.~Bush, ``Securing {BGP} and {SIDR},'' \emph{IETF Journal}, vol.~7, no.~1, 2011.

\bibitem{AP2Vec}
T.~Shapira and Y.~Shavitt, ``{AP2Vec}: an unsupervised approach for {BGP} hijacking detection,'' \emph{IEEE Transactions on Network and Service Management}, 2022.

\bibitem{Gao-VF}
L.~Gao, ``On inferring autonomous system relationships in the internet,'' \emph{{IEEE/ACM} Transactions on Networking}, vol.~9, no.~6, pp. 733--745, 2001.

\bibitem{RV}
U.~of~Oregon Advanced Network Technology~Center, ``Route views project,'' \url{http://www.routeviews.org/}.

\bibitem{DeepBGP-2020}
M.~Bahnasy, F.~Li, S.~Xiao, and X.~Cheng, ``{DeepBGP}: A machine learning approach for {BGP} configuration synthesis,'' in \emph{SIGCOMM Workshop on Network Meets AI \& ML (NetAI'20)}, Aug. 2020, pp. 48--55.

\end{thebibliography}

\end{document}